\documentclass{emulateapj} 

\shorttitle{Magnetic field in NGC\,4736}
\shortauthors{K.~T. Chy\.zy  and R.~J. Buta}

\begin{document}

\title{Discovery of a strong spiral magnetic field crossing the 
inner pseudoring of NGC\,4736}

\author{Krzysztof T. Chy\.zy\altaffilmark{1} and Ronald J. Buta\altaffilmark{2}}
\altaffiltext{1}{
Astronomical Observatory, Jagiellonian University,
30-244 Krak\'ow, Poland; chris@oa.uj.edu.pl
}
\altaffiltext{2}{
Department of Physics and Astronomy, University of Alabama,
Tuscaloosa, AL 35487, USA}

\begin{abstract}
We report the discovery of a coherent magnetic {\em spiral} structure within 
the nearby {\em ringed} Sab galaxy NGC\,4736. High sensitivity radio 
polarimetric data obtained with the VLA at 8.46\,GHz and 4.86\,GHz show a 
distinct ring of {\em total} radio emission precisely corresponding to the 
bright inner pseudoring visible in other wavelengths. However, unlike the total 
radio emission, the {\em polarized} radio emission reveals a clear pattern of 
ordered magnetic field of spiral shape, emerging from the galactic centre. The 
magnetic vectors do not follow the tightly-wrapped spiral arms that 
characterize the inner pseudoring, but instead cross the ring with a constant 
and large pitch angle of about $35^{\circ}$. The ordered field is thus not locally 
adjusted to the pattern of star-formation activity, unlike what is usually 
observed in grand-design spirals. The observed asymmetric distribution of 
Faraday rotation suggests the possible action of a large-scale MHD dynamo. 
The strong magnetic total and regular field within the ring (up to $30\mu$G 
and $13\mu$G, respectively) indicates that a highly efficient process of 
magnetic field amplification is under way, probably related to  
secular evolutionary processes in the galaxy.

\end{abstract}
\keywords{galaxies: individual (NGC 4736) --- galaxies: magnetic fields --- MHD}

\begin{figure*}
\centering
\includegraphics[width=8.9cm]{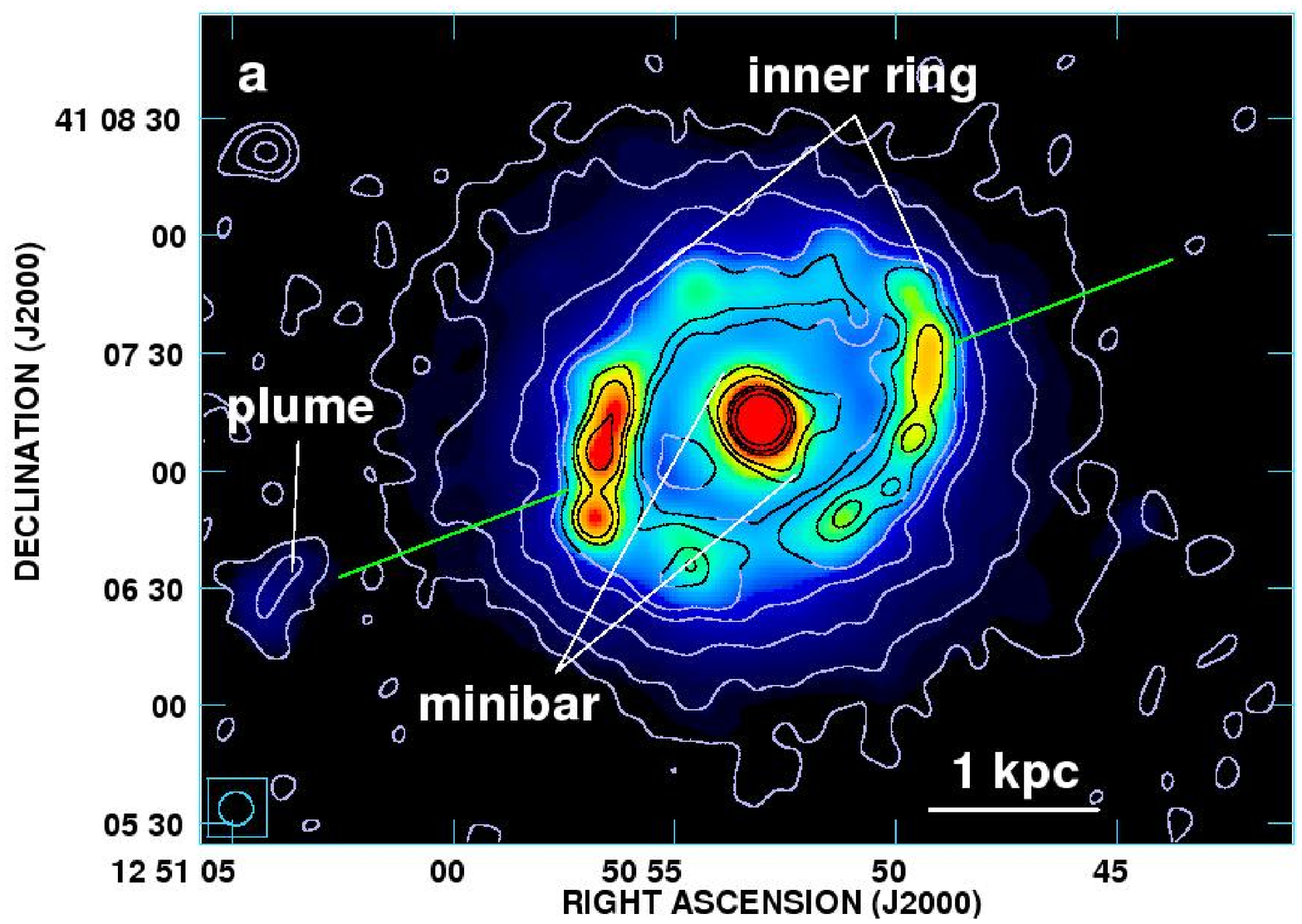}
\includegraphics[width=8.9cm]{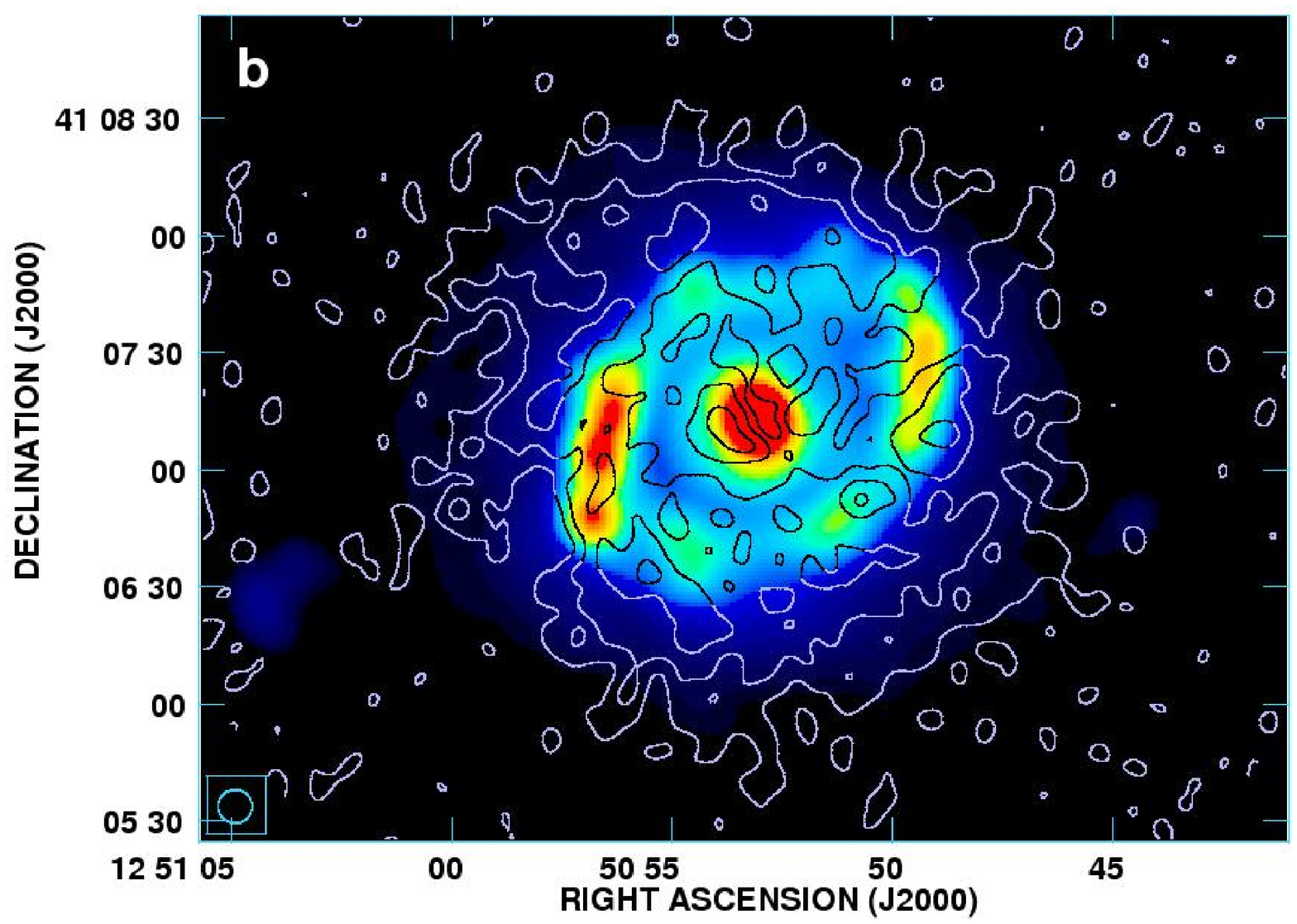}
\caption{(a) The total radio intensity contour map of NGC\,4736 at 8.46\,GHz
and 8\rlap{.}$^{\prime\prime}$6x8\rlap{.}$^{\prime\prime}$6 resolution 
(from combined VLA and Effelsberg data) superimposed upon the infrared 
$24\,\mu$m image \citep[in colours, from Spitzer survey of SINGS galaxies:]
[]{kennicutt03}. The contours are at 28, 72, 176, 320, 440 800, 1040, 1600 
$\mu$Jy/beam. (b) The radio polarized intensity at 8.46\,GHz in contours and 
the infrared map in colours. The contours are at 21, 42, 70, 98 $\mu$Jy/beam. 
The galaxy is inclined by $35^{\circ}$ \citep{buta88} and the green 
line in (a) denotes its major axis. The inner ($\underline{\rm r}$s) ring 
and the central minibar are indicated, whereas the faint, outer (R) ring 
is out of the figure and also invisible in the radio data.}
\label{f:total}
\end{figure*}

\begin{figure}
\centering
\includegraphics[width=8.9cm]{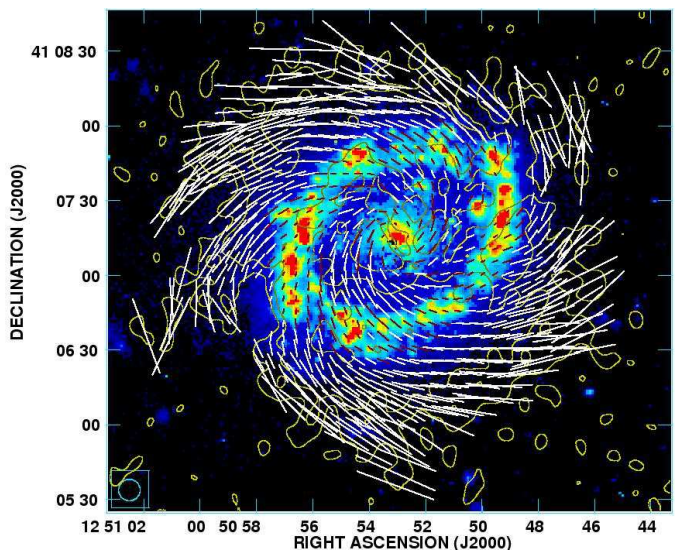}
\caption{The polarized intensity contour map of NGC\,4736 at 8.46\,GHz and 
8\rlap{.}$^{\prime\prime}$6x8\rlap{.}$^{\prime\prime}$6 resolution with 
observed magnetic field vectors of the polarization degree overlaid upon the 
H$\alpha$ image \citep[from][]{knapen03}. The contours are at 21, 42, 84 
$\mu$Jy/beam area. The vector of $10\arcsec$ corresponds to the polarization 
degree of 25\%.}
\label{f:3cmpi}
\end{figure}

\begin{figure*}
\centering
\includegraphics[width=15cm]{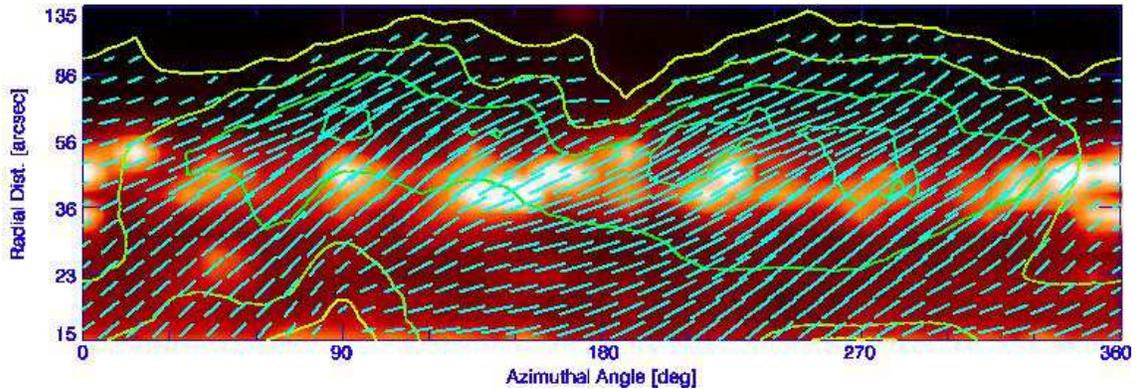}
\caption{A diagram of the de-projected regular magnetic field vectors in
the plane of NGC\,4736 at 8.46\,GHz (without correction for Faraday effects). 
The galactocentric azimuthal angle is measured counterclockwise from the 
northern tip of the major axis (P.A.=295$^{\circ}$). The vector's length is 
proportional to the polarized intensity, also presented in contours. The 
H$\alpha$ image is shown in colours.}
\label{f:phase}
\end{figure*}

\begin{figure}
\centering
\includegraphics[width=8.5cm]{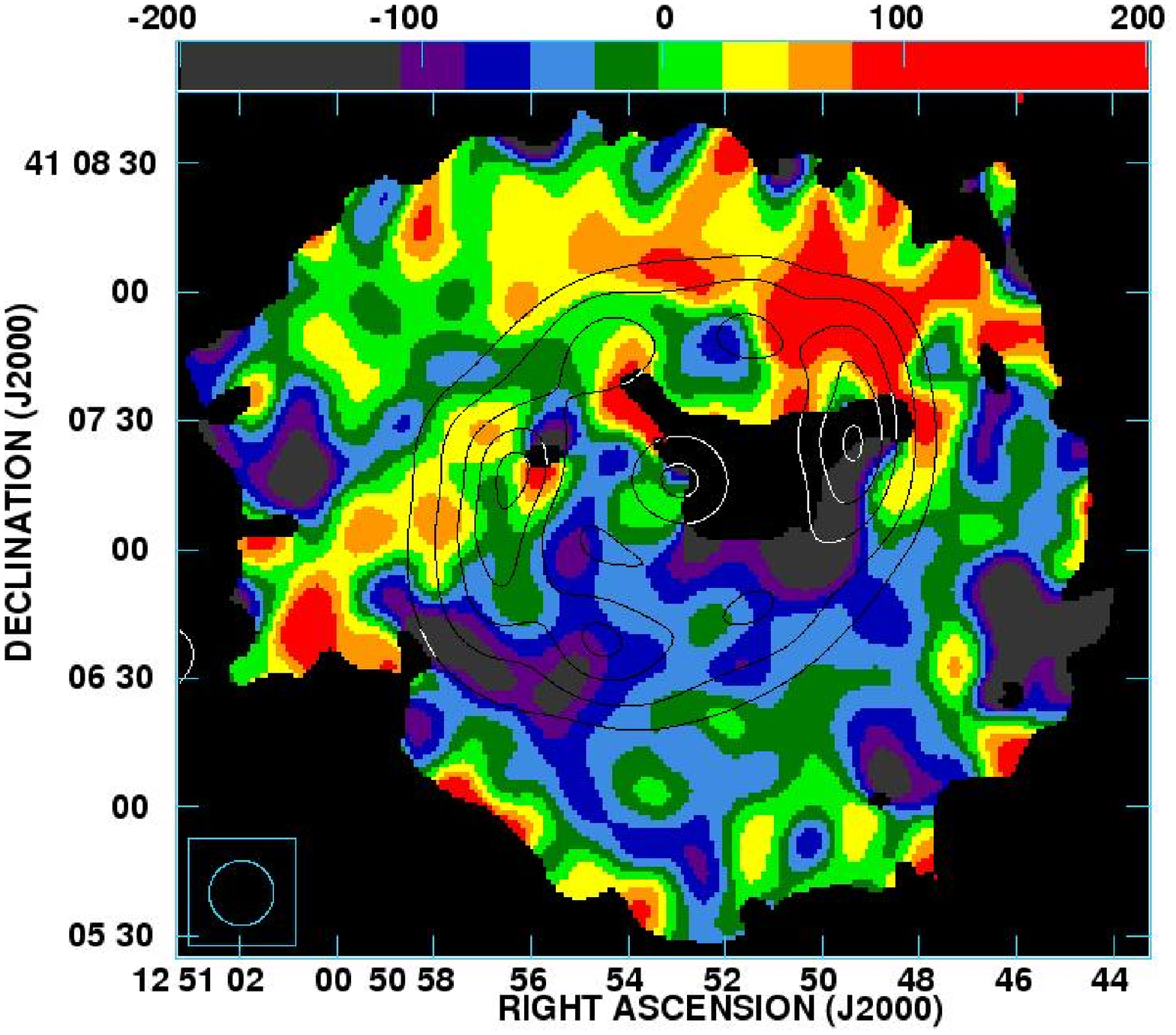} 
\caption{Faraday rotation measure distribution (in colours) of NGC\,4254,
at $15\arcsec$ resolution, computed from 8.46\,GHz and 4.86\,GHz data, 
in rad~m$^{-2}$, with contours of H$\alpha$ emission.}
\label{f:rm}
\end{figure}

\section{Introduction}
\label{s:intro}

During the last decade, galactic magnetic fields have been observed in detail 
in a number of galaxies, from the grand-design and barred spirals to the dwarf 
irregulars \citep{beck05,chyzy00}. However, the role of magnetic fields in the 
dynamical evolution of galaxies and their interstellar medium (ISM), and the 
conditions and efficiency of the magnetic field generation process, are still 
not well understood. It is also unclear how the magnetohydrodynamic (MHD) 
dynamo interplays with gas flows associated with spiral density waves 
\citep{elstner05,shukurov05}. In grand-design spirals the magnetic field pitch 
angle adjusts to the high concentration of star formation in the spiral arms 
as e.g. in NGC\,6946 \citep{beck07}, or in perturbed NGC\,4254 \citep{chyzy07}. 
But in flocculent galaxies, the possibly dynamo-generated magnetic field shows 
a large pitch angle ($25^{\circ}$-$30^{\circ}$), that sometimes agrees with optical 
armlets, and sometimes not \citep{knapik00}. 

In contrast to grand-design spirals and many barred galaxies dominated by the 
spiral arms there are some early-type galaxies classified by de Vaucouleurs as 
``SA(r)" which are purely dominated by rings. Unlike spiral arms, galactic 
rings are believed to be formed by resonant accumulation of gas in the 
non-axisymmetric potential of galaxies \citep{buta96,buta07}. 
Under the continuous action of gravity torques gas is driven into resonance 
regions whose shape and orientation can be determined by the properties of 
periodic orbits \citep[e.g.][]{regan04}. 
Such galaxies of ringed morphology and resonant dynamics can 
be excellent targets to address the above questions of galactic magnetism. 
Without strong spiral density waves and associated streaming motions they can 
complete our knowledge about the dynamo process in galaxies showing no 
grand-design spiral patterns.

Motivated by these goals we performed radio polarimetric observations of 
NGC\,4736 (M\,94), the nearest and largest galaxy in the sky to manifest 
a strong ringed morphology. Classified as type 
(R)S$\underline{\rm A}$B($\underline{\rm r}$s)ab by \citet{buta07}, it shows 
two rings. The outer ring (R) is a weak, diffuse, and smooth feature. 
But the inner pseudoring ($\underline{\rm r}$s) is a well-defined zone of 
intense active star formation visible in optical, H$\alpha$, UV emission 
\citep{waller01}, and in infrared (Fig.~\ref{f:total}) and CO maps 
\citep{wong00}. The inner ring extends to a radial distance of about 
$47\arcsec$ (1.1 kpc)\footnote{We adopt a distance of 4.66\,Mpc 
\citep{karachentsev04}} around the galaxy (LINER) nucleus. NGC\,4736 is not 
completely axisymmetric but has a large-scale, broad oval that may affect 
its internal dynamics \citep{kormendy04}. Hydrodynamic simulations suggest that 
the rotating oval is indeed responsible for the rings' formation 
\citep{mulder96}. The nuclear minibar, well visible in CO emission as a 
central $20\arcsec$ extension perpendicular to the galaxy major axis, may also 
participate in the ring formation process \citep{wong00}. 

\section{Radio observations and results}
\label{s:obs}

We observed NGC\,4736 between 8 April and 16 May 2007 at 8.46\,GHz and 
4.86\,GHz with the Very Large Array (VLA\footnote{The VLA of the NRAO is 
operated by Associated Universities, Inc., under cooperative agreement with 
the NSF}) in its D-array. The data obtained were reduced, calibrated, and self-calibrated 
with the standard AIPS package. The allocated observing time of 33\,h enabled 
us to get high sensitivity radio maps: in the polarized intensity we reached 
an rms noise level of $6\,\mu$Jy/beam at 8.46\,GHz in an 
8\rlap{.}$^{\prime\prime}$6x8\rlap{.}$^{\prime\prime}$6 resolution map, 
and likewise $12\,\mu$Jy/beam at 4.86\,GHz in a $15\arcsec$x$15\arcsec$ 
resolution map.

To avoid the missing-spacing problem, we also made radio polarimetric 
observations of NGC\,4736 with the Effelsberg 100-m telescope at frequencies of 
8.35\,GHz and 4.85\,GHz. The data reduction and merging with the 
interferometric data were performed in a manner similar to NGC\,4254 
\citep{chyzy07}. The combined maps in I, Q, and U Stokes parameters at both 
frequencies were then used to construct maps of polarized intensity, 
polarization position angle, position angle of observed magnetic vectors 
(observed $E$--vectors rotated by $90\degr$), and Faraday rotation measure. 

The observed total radio emission of NGC\,4736 at 8.46\,GHz is clearly 
dominated by the galaxy's bright inner pseudoring (Fig.~\ref{f:total}a) 
and resembles the distributions of infrared, H$\alpha$, and UV emission 
\citep[e.g.][]{waller01}. All pronounced radio features in the ring correspond 
to the enhanced signal in the mid-infrared (in colours in Fig.~\ref{f:total}a) 
and must result from an intense star-formation process providing dust heating 
and strong radio thermal and nonthermal emission. The radio contours in the 
galaxy's bright bulge region (within $20\arcsec$ radius) are slightly elongated 
in position angle P.A.$\approx$30$^{\circ}$. They likely correspond to the nuclear 
minibar seen in optical and CO images (Sect.~\ref{s:intro}). Outside the 
galaxy's bright radio disk, weak radio emission is detected from a star-forming 
plume (Fig.~\ref{f:total}a), being another feature of the galaxy's resonant 
dynamics \citep{waller01}.

The polarized radio emission of NGC\,4736 at 8.46\,GHz reveals a dramatically 
different morphology (Fig.~\ref{f:total}b). It does not clearly correspond 
either to the inner pseudoring in the infrared emission or to the 
distribution of total radio emission. The degree of polarization is slightly 
lower in the ring (about $10\%\pm1\%$ on average) than in its close vicinity 
($15\%\pm1\%$), and rises to about $40\%$ at the disk edges. The observed 
vectors of regular magnetic field (Fig.~\ref{f:3cmpi}) are organized into a 
very clear spiral pattern with two broad magnetic arms. Surprisingly, the inner 
ring hardly influences the magnetic vectors: they seem to cross the 
star-forming regions without any change of their orientation. This is 
opposite to what is observed in grand-design spiral galaxies 
(Sect.~\ref{s:intro}), where the magnetic field typically follows 
a nearby spiral density wave.

The revealed spiral magnetic pattern at 8.46\,GHz, is fully confirmed at 
4.86\,GHz. The observed similar orientation of magnetic field vectors at both 
radio frequencies indicates only small Faraday rotation effects in this galaxy 
(see below). Hence, the magnetic vectors presented in Figure~\ref{f:3cmpi} 
give almost precisely the intrinsic direction of the projected magnetic field 
(within 7$^{\circ}$) in most of the galactic regions.

\section{Pure dynamo action?}
\label{s:dynamo}

The observed spiral structure of the magnetic field in NGC\,4736 contradicts 
the main feature of its optical morphology: the starbursting inner pseudoring. 
To investigate the exact pattern of regular magnetic field without projection 
effects, we constructed a phase diagram (Fig.~\ref{f:phase}) of magnetic 
field vectors along the azimuthal angle in the galaxy plane versus the natural 
logarithm of the galactocentric radius. It confirms that the magnetic vectors 
cross the H$\alpha$ emitting ring (which constitutes a horizontal structure in 
Fig.~\ref{f:phase}) without changing their large pitch angle of 35$^{\circ}\pm5^o$. 
The two broad magnetic spiral arms (Sect.~\ref{s:obs}) clearly emerge from 
close to the galactic centre, at azimuths of about $0^{\circ}$ and $180^{\circ}$. Inside the 
inner ring, around azimuths of $120^{\circ}$ and $300^{\circ}$, the magnetic pitch angle 
attains smaller values, from $0^{\circ}$ to $20^{\circ}$, which may result from gas-flows 
around the central minibar (Sect.~\ref{s:intro}). Despite this, the observed 
pattern of regular field in NGC\,4736 seems to be the most coherent one 
observed so far in spiral galaxies \citep[cf.][]{beck05}. 

The comparison of the magnetic pattern in NGC\,4736 with the Hubble Space 
Telescope and other filtered optical images indicates further disagreement of 
the regular magnetic field with other galactic structures, e.g. the prominent, 
almost circular, long dust lane west of the centre \citep{waller01}, or spiral 
dust armlets (probably of acoustic origin) in the central part of the galaxy 
\citep{elmegreen02}. Outside the ring, the relation of magnetic vectors with 
optical, flocculent features \citep{waller01} is ambiguous, and contrary to the 
magnetic structure the optical features do not continue inside the ring. The 
kinematics of CO- and HI-emitting gas near the ring is well-described by pure 
circular differential rotation with a velocity of about 200\,km\,s$^{-1}$ and 
small residuals, typically less than about 10\,km\,s$^{-1}$ \citep{wong00}. 
This assures that galactic shearing motions in the vicinity of the inner ring 
are strong. Hence, the observed magnetic spiral could result from a {\em pure} 
MHD dynamo action that develops without support from spiral density waves 
(Sect.~\ref{s:intro}).

The strongest observational test for the origin of the galactic magnetic field 
is the distribution of Faraday rotation measure (RM), which is sensitive to 
the sense of direction of the magnetic field. Magnetic fields produced locally 
by ejections from stars or by small-scale MHD dynamos, compressed in shocks or 
stretched by gas shearing flows, yield random fields and incoherent (changing 
sign) RM patterns. Only the large-scale dynamo can induce unidirectional 
magnetic field and produce a coherent RM pattern on the galactic-scale. 
The typical RM values observed in NGC\,4736 (Fig.~\ref{f:rm}) 
are small, about $\pm$50\,rad\,m$^{-2}$, reaching locally 
$|{\rm RM}|>100$\,rad\,m$^{-2}$. As the galaxy is located at high Galactic 
latitude ($76^{\circ}$) the influence of the Milky Way on observed RM could be 
omitted. Globally, NGC\,4736 shows a large area of statistically positive RM in 
the NW part of the galaxy and negative RM in the SE one. This gives a strong 
argument for a large-scale MHD dynamo working in this object.

\begin{figure}[t]
\centering
\includegraphics[width=2.8cm]{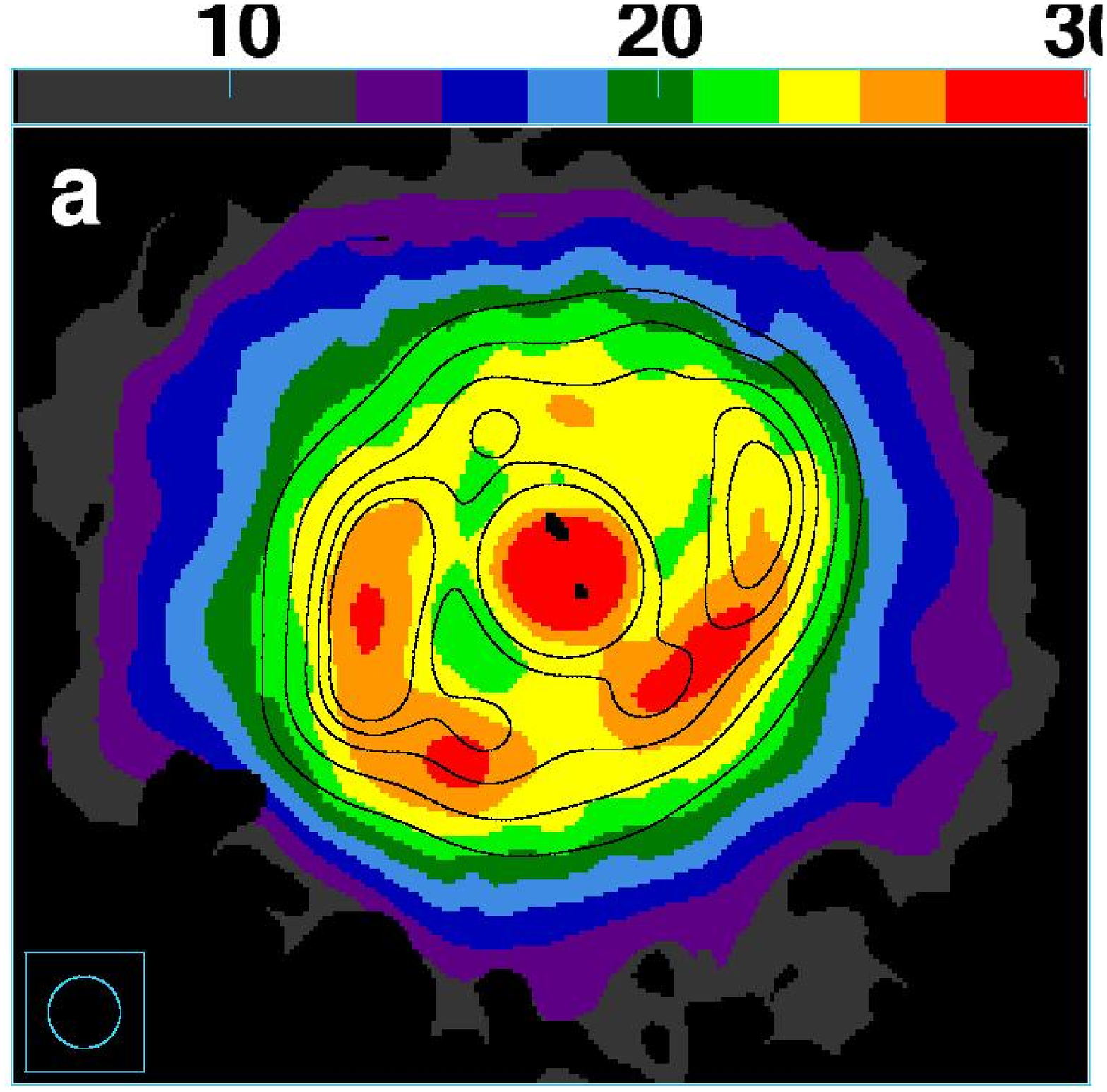}
\includegraphics[width=2.8cm]{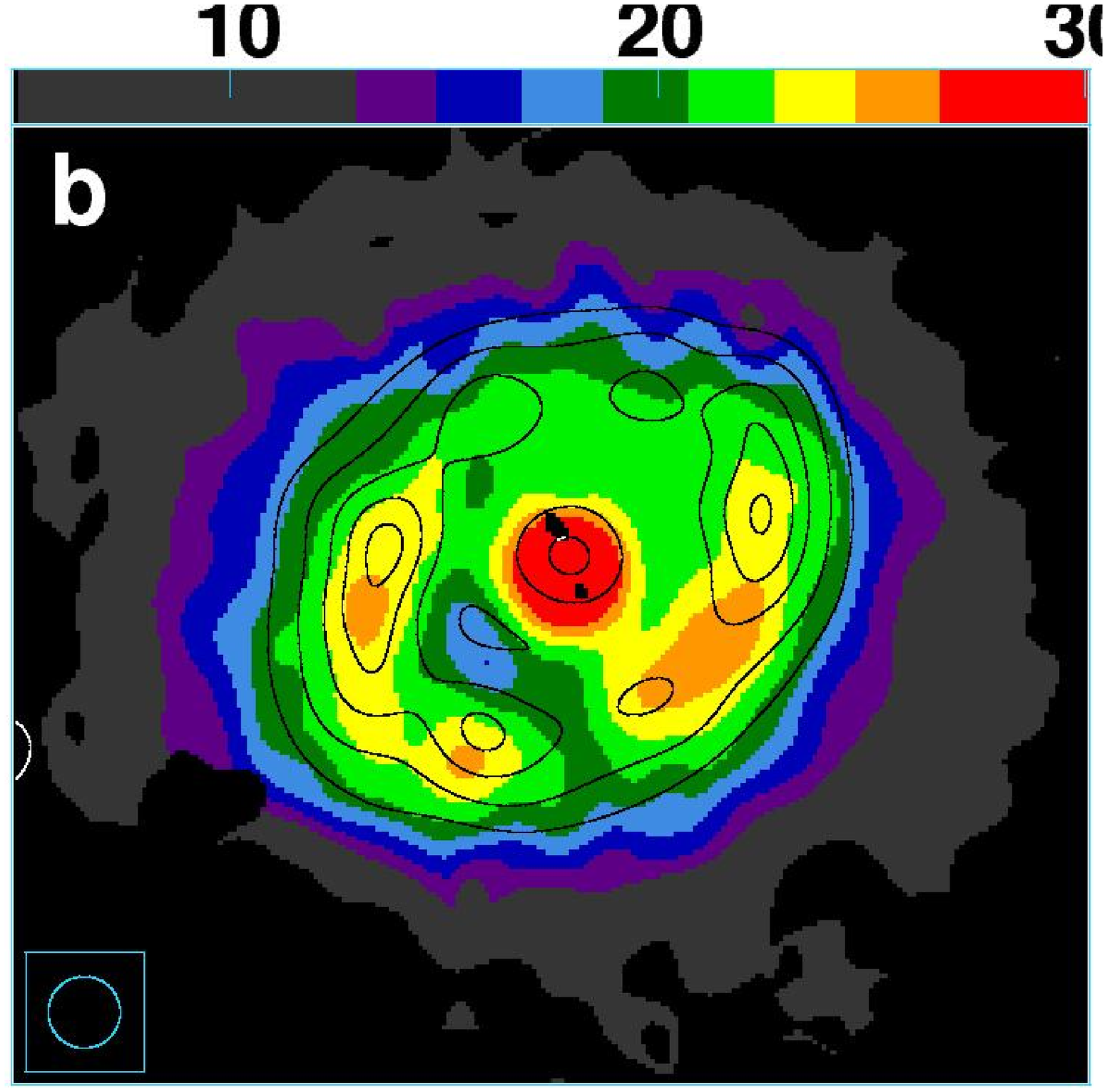}
\includegraphics[width=2.8cm]{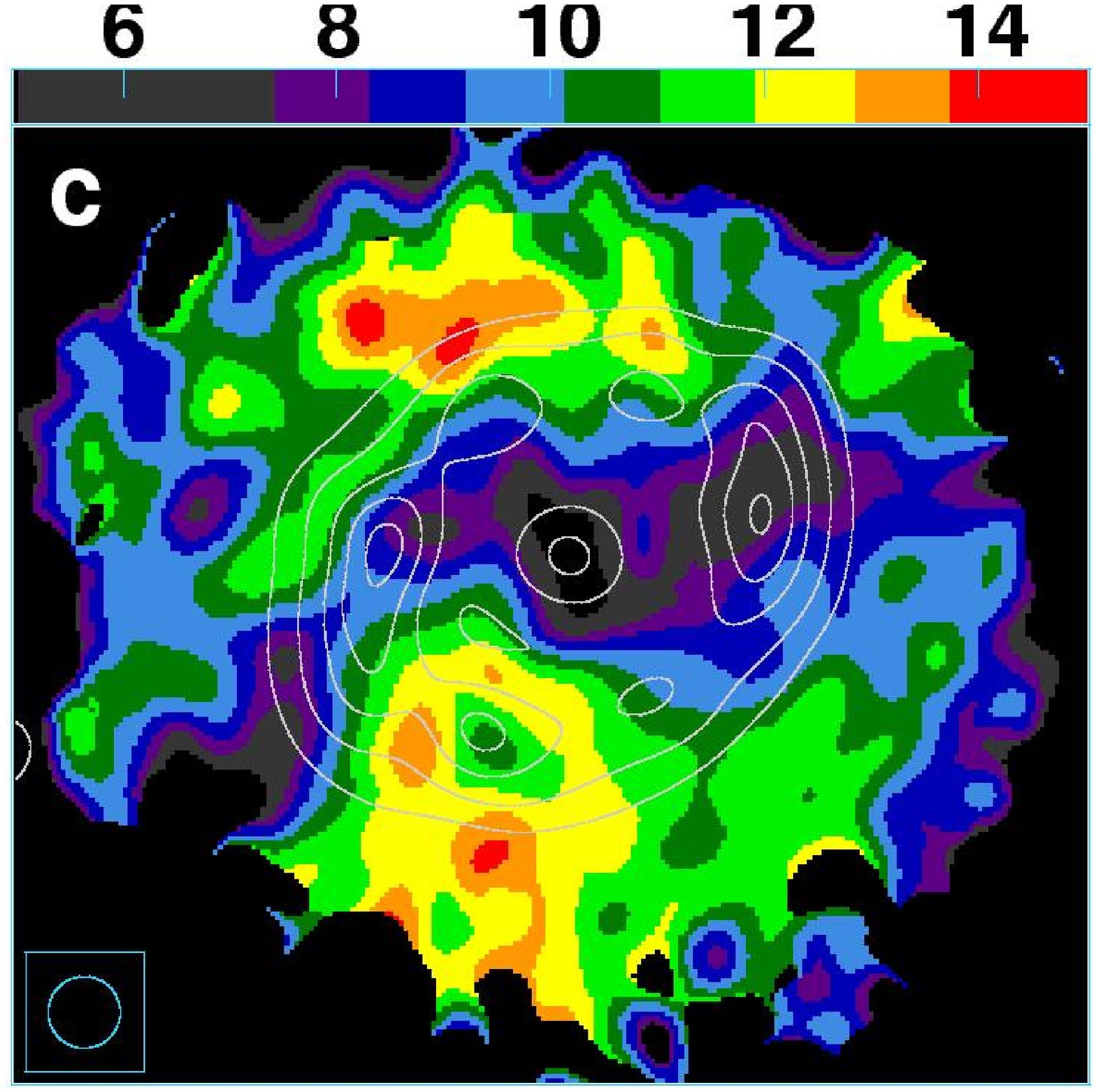}
\caption{The magnetic maps of NGC\,4736: (a) total, (b) random, and (c) regular 
magnetic field strength (in $\mu$G) in colours with contours of infrared 
24\,$\mu$m (a) and H$\alpha$ emission (b, c).}
\label{f:mmaps}
\end{figure}

From maps of total and polarized emission and the direction of magnetic 
field vectors corrected for Faraday rotation we derive ``magnetic maps'' 
\citep{chyzy08} - the strength of total, random, and regular magnetic field 
throughout the galaxy plane, corrected for projection effects. In calculations 
we assume equipartition between the energy of the magnetic field and cosmic 
rays (CR) with an $E$=300 MeV cutoff in the CR proton spectrum, the energy 
ratio $k$=100 of CR protons and electrons, and an unprojected synchrotron disk 
thickness of $L$=500 pc. The thermal emission is separated from the observed 
radio intensity assuming a non-thermal spectral index of 0.9.
The total magnetic field (Fig.~\ref{f:mmaps}a) is strongest in the galactic 
centre and in the starbursting ring where its strength varies from 18\,$\mu$G 
to even 30\,$\mu$G. The field is dominated by the random component 
(Fig.~\ref{f:mmaps}b) which roughly correlates with star-forming regions of 
H$\alpha$ emission, as also seen in other galaxies. Contrary to the random 
component the regular magnetic field (Fig.~\ref{f:mmaps}c) is strongest in the 
southern part of the ring and in the regions outside the ring, reaching locally 
13\,$\mu$G. These local values are similar to the largest ones found in the 
non-barred late-type spirals \citep{beck05}. 

The mean total and regular field within the radio extent of 
NGC\,4736 (of $\approx$3\rlap{.}$^{\prime}$5) is 17\,$\mu$G and 10\,$\mu$G, 
respectively. These mean values are much larger than the averages found for 
spiral galaxies of various types \citep[9\,$\mu$G and 4\,$\mu$G, respectively;]
[]{beck05}. Our estimates of magnetic field strengths for NGC\,4736 depend 
somewhat on assumptions given above. If instead of equipartition, pressure 
equilibrium is assumed (yielding minimum stable magnetic field), its strength 
is lowered by about 25\%. The magnetic field may vary by the same amount if 
the values of $E$, $k$, and $L$ are varied independenty by 50\%. We thus 
conclude that the MHD dynamo in NGC\,4736 must be very efficient.
 
\section{Origin of magnetic field}
\label{s:discussion}

We have shown that the spiral pattern of regular magnetic field lines in 
NGC\,4736 is not coupled to any gas distribution features or recognized gas 
flows and can be of pure dynamo origin, unlike those of grand-design spirals 
(Sect.~\ref{s:dynamo}). The map of Faraday rotation (Fig.~\ref{f:rm}) confirms 
that some large-scale dynamo must be at work. If an evolving 
freely dynamo, without strong support from spiral density waves, produces a spiral 
field like that seen in NGC\,4736, then such a dynamo could also explain 
magnetic spiral patterns found in several flocculent galaxies lacking obvious 
spiral density waves (Sect.~\ref{s:intro}). 

Observations of NGC\,4736 can also help in understanding the spiral magnetic 
field revealed in the area of the circumnuclear ring of NGC 1097 
\citep{beck99}, a barred spiral galaxy, strongly interacting with a nearby 
companion \citep[see][]{buta07}. According to a recent interpretation 
\citep{prieto05}, the nuclear magnetic spiral in this galaxy is coupled to 
extraplanar gas. Such an explanation does not seem 
likely for the spiral magnetic field in NGC\,4736 due to the very coherent 
magnetic pattern and small depolarization effects in the ring. Also, NGC\,4736 
is a relatively isolated galaxy, suggesting that internal dynamics play a more 
important role. The spiral fields around the star-forming rings in both 
galaxies could therefore be of different origins. But the alternative scenario 
that the structure of the magnetic field in the center of NGC\,1097 may be 
dominated by a pure dynamo action (as in NGC\,4736) could also be valid. 
Indeed, unlike the bar region and beyond, strong density waves in NGC\,1097 are 
largely absent inside the circumnuclear ring.

The high efficiency of the dynamo amplification process in NGC\,4736 is 
surprising for such an early-type object, because in another early-type 
spiral, the Sombrero galaxy NGC\,4594, the magnetic field is four times weaker 
and reaches locally 6\,$\mu$G in the total and only 3\,$\mu$G in the regular 
component \citep{krause06}. The cause for this discrepancy may lie in different 
galaxy evolutions. While NGC\,4594 has a classical bulge, NGC\,4736 contains 
a prototypical pseudobulge \citep{kormendy04}, with different S\'ersic index 
$n$ in the luminosity profile ${\rm log}I\propto r^{1/n}$, and a larger 
flattening, revealing much more regular (disky) rotation. Thus the secular 
evolution of NGC\,4736, which also led to formation of the starbursting 
pseudoring, might affect the magnetic field generation and evolution, and 
account for the observed difference between the two galaxies. 

The reported large pitch angle of the regular magnetic field in NGC\,4736 
($35^{\circ}$, Fig.~\ref{f:phase}) demands special conditions for the dynamo to be 
set-up. According to the MHD simulation by \citet{elstner00}, in typical spiral 
galaxies the large pitch angles ($>20^{\circ}$) are supported by spiral density 
waves. These are, however, weak in NGC\,4736, and in the region of its inner 
ring they become almost circular. For $\alpha\Omega$-dynamos with 
$\alpha$-quenching one has a dependence of the magnetic pitch angle $p$ on the 
turbulent diffusivity $\eta_T$, the disk scale height $h$, the turnover radius 
$r_\Omega$ in the rotation curve, and the rotational velocity $v$ 
\citep{elstner05}: ${\rm tan}\,p \approx 3\,r_\Omega \, \eta_T/(v \, h^2)$.
With $r_\Omega$=440\,pc and $v$=200\,km\,s$^{-1}$ \citep[estimated from the 
CO and HI rotation curve;][]{wong00}, and with typically used values for 
late-type spirals of $\eta_T$=1\,kpc\,km\,s$^{-1}$ and $h$=500\,pc we get 
$p$=$2\degr$, much less than the observed value. To produce larger $p$ by a 
pure dynamo process, a larger magnetic diffusivity or thinner gaseous disk is 
necessary. The H$\alpha$ and synchrotron scale heights of the ring are 370\,pc 
and 480\,pc, respectively (c.f. Fig.~\ref{f:total}). Thus, the disk scale 
height is unlikely to be $\le$300\,pc which leads to a needed turbulent 
diffusion of at least $10$\,kpc\,km\,s$^{-1}$. Such conditions could likely be 
achieved in the violently star-forming inner ring with the support of e.g. a 
strong diffusion of CRs, as in the fast Parker's cosmic rays-driven dynamo 
\citep{hanasz04}. The fact that the radio polarized emission is 
so highly restricted to the vicinity of the starbursting ring 
(Fig.~\ref{f:3cmpi}) provides another stringent constraint on the dynamo 
action. Thus, NGC\,4736 is an excellent object to test the most recent concepts 
of MHD dynamos.

In this report we demonstrated that the early-type galaxy NGC\,4736 has highly 
symmetric and strong magnetic spiral fields, not clearly associated with the 
shape of the distribution of star-forming regions or with spiral density waves. 
The detailed processes which cause the dynamo in NGC\,4736 to run so 
efficiently and enable the generated regular field to ignore the galaxy's 
ringed morphology remain to be determined.

\acknowledgments
We thank R. Beck for help in gathering the Effelsberg data and R. Beck, 
D. Elstner, M. Krause, K. Otmianowska-Mazur, M. Urbanik, and the referee for 
helpful comments. KTC acknowledges the support of MNiSW grant 
2693/H03/2006/31 and RJB the support of NSF grant AST 050-7140.


\begin{thebibliography}{}

\bibitem[Beck et al.(1999)]{beck99} 
  Beck, R., Ehle, M., Shoutenkov, V., Shukurov, A., \& Sokoloff, 
  D. 1999, \nat, 397, 324 
\bibitem[Beck(2005)]{beck05} 
  Beck, R. 2005, In: Cosmic Magnetic Fields, eds.
  R.\,Wielebinski \& R.\,Beck (Heidelberg: Springer), p.~41
\bibitem[Beck(2007)]{beck07} 
  Beck, R. 2007, A\&A, 470, 539
\bibitem[Buta(1988)]{buta88} 
  Buta, R. 1988, ApJS, 66, 233
\bibitem[Buta \& Combes(1996)]{buta96} 
  Buta, R., \& Combes, F. 1996, Fund. Cosmic Phys., 17, 95
\bibitem[Buta et al.(2007)]{buta07} 
  Buta, R., Corwin, H. G., \& Odewahn, S. C. 2007, The de Vaucouleurs Atlas
  of Galaxies, Cambridge University Press
\bibitem[Chy{\.z}y et al.(2000)]{chyzy00} 
  Chy{\.z}y, K.~T., Beck, R., Kohle, S., Klein, U., \& Urbanik, M. 2000, A\&A, 355, 128 
\bibitem[Chy\.zy et al.(2007)]{chyzy07} 
  Chy\.zy, K. T., Ehle, M., \& Beck, R. 2007, A\&A, 474, 415
\bibitem[Chy\.zy(2008)]{chyzy08} 
  Chy\.zy, K.T. 2008, A\&A, in press , astro-ph 0712.4175
\bibitem[Elmegreen et al.(2002)]{elmegreen02}
  Elmegreen, D.M., Elmegreen, B.G., \& Eberwein, K.S. 2002, ApJ, 564, 234
\bibitem[Elstner et al.(2000)]{elstner00} 
  Elstner D., Otmianowska-Mazur, K., von Linden, S., \& Urbanik, M. 
  2000, A\&A, 357, 129
\bibitem[Elstner(2005)]{elstner05} 
  Elstner D. 2005, in {\em The Magn. Plasma in Galaxy Evolution}, Krakow 2005, p117
\bibitem[Hanasz et al.(2004)]{hanasz04} 
  Hanasz, M., Kowal, G., Otmianowska-Mazur, K., \& Lesch, H. 2004, \apjl, 605, L33 
\bibitem[Karachentsev et al.(2004)]{karachentsev04} 
  Karachentsev, I. D., Karachentseva, V. E., Huchtmeier,W. K., \& Makarov, 
  D. I. 2004, AJ, 127, 2031
\bibitem[Kennicutt et al.(2003)]{kennicutt03} 
  Kennicutt, R. C., Armus, L., Bendo, G. et al. 2003, PASP, 115, 928
\bibitem[Knapen et al.(2003)]{knapen03} 
  Knapen, J. H., de Jong, R. S., Stedman, S., \& Bramich, D.M. 2003, 
  MNRAS, 344, 527
\bibitem[Knapik et al.(2000)]{knapik00}
  Knapik, J., Soida, M., Dettmar, R.-J., Beck, R., \& Urbanik, M. 2000,
  A\&A, 362, 910  
\bibitem[Kormendy \& Kennicutt(2004)]{kormendy04} 
  Kormendy, J., \& Kennicutt, R.C. 2004, \araa, 42, 603
\bibitem[Krause et al.(2006)]{krause06} 
  Krause, M., Wielebinski, R., \& Dumke, M. 2006, A\&A, 448, 133
\bibitem[Mulder \& Combes(1996)]{mulder96} 
  Mulder, P.S. \& Combes, F. 1996, A\&A, 313, 723
\bibitem[Prieto et al.(2005)]{prieto05} 
  Prieto, M.~A., Maciejewski, W., \& Reunanen, J. 2005, \aj, 130, 1472 
\bibitem[Regan \& Teuben(2004)]{regan04} 
  Regan, M. \& Teuben, P. 2004, \apj, 600, 595
\bibitem[Shukurov(2005)]{shukurov05}
  Shukurov, A. 2005, In: Cosmic Magnetic Fields, eds.
  R.\,Wielebinski \& R.\,Beck (Heidelberg: Springer), p.~111
\bibitem[Waller et al.(2001)]{waller01} 
  Waller, H.W., Fanelli, M.N., Keel, W.C., et al. 2001, AJ, 121, 1395
\bibitem[Wong \& Blitz(2000)]{wong00} 
  Wong, T., \& Blitz, L. 2000, ApJ, 540, 771
\end{thebibliography}
\end{document}